\documentclass[aps,showpacs,nofootinbib]{revtex4}

\usepackage{graphicx}
\usepackage{amsmath}
\usepackage{amssymb}

\newcommand{\be}{\begin{equation}}
\newcommand{\ee}{\end{equation}}
\newcommand{\ba}{\begin{eqnarray}}
\newcommand{\ea}{\end{eqnarray}}

\def\lb{\label}

\begin{document}

\title{Relativistic field theories in a magnetic background as 
noncommutative field theories}

\author{E.V. Gorbar}
  \email{egorbar@uwo.ca}
  \altaffiliation[On leave from ]{
       Bogolyubov Institute for Theoretical Physics,
       03143, Kiev, Ukraine
}

\author{V.A. Miransky}
  \email{vmiransk@uwo.ca}
   \altaffiliation[On leave from ]{
       Bogolyubov Institute for Theoretical Physics,
       03143, Kiev, Ukraine
}
   
\affiliation{
Department of Applied Mathematics, University of Western
Ontario, London, Ontario N6A 5B7, Canada
}

\date{\today}

\begin{abstract}
We study the connection of the dynamics in relativistic field 
theories in a strong magnetic field
with the dynamics of noncommutative field theories (NCFT). 
As an example, the 
Nambu-Jona-Lasinio models in spatial dimensions  
$d \geq 2$ are considered. We show
that this connection is rather sophisticated. In fact,
the corresponding NCFT are different from the conventional
ones considered in the literature. In particular,
the UV/IR mixing is absent in these theories. The reason of that
is an inner structure (i.e., dynamical form-factors) of neutral
composites which plays an important role in providing
consistency of the NCFT.
An especially interesting case is that for
a magnetic field configuration with the maximal number 
of independent nonzero tensor components. In that case,
we show that 
the NCFT are finite for even $d$ and their dynamics is
quasi-(1+1)-dimensional
for
odd $d$. For even $d$, the NCFT
describe a confinement dynamics of charged particles. The
difference between the dynamics in strong magnetic backgrounds  
in field theories and that in string theories is briefly discussed.
\end{abstract}


\maketitle

\section{Introduction}
\label{1}

Recently, there has been a considerable interest in
noncommutative field theories (NCFT) (for reviews, see Ref.
\cite{DNS}).
Besides being interesting in themselves, 
noncommutative theories mimic certain dynamics
in quantum mechanical models \cite{DJT,BS}, nonrelativistic field 
systems \cite{IKS,CTZ},
nonrelativistic magnetohydrodynamical field theory \cite{GJPP},
and string theories 
\cite{strings,SW}. In
particular, NCFT are intimately related to the dynamics in
quantum mechanical and nonrelativistic field 
systems in a strong magnetic 
field 
\cite{DJT,BS,IKS,CTZ,GJPP} 
and, in the case of open strings
attached to $D$-branes, to the dynamics in string theories in
magnetic backgrounds \cite{BS,SW,Sheikh}.

In this paper, we study 
the connection between the dynamics 
in relativistic field theories
in a strong magnetic field  
and that in NCFT.
Our main conclusion
is that although field theories in the regime with the
lowest Landau level (LLL) dominance
indeed determine a class
of NCFT, these NCFT are different from
the conventional ones considered in the literature.
In particular,
the UV/IR mixing, taking place in the
conventional NCFT \cite{MRS}, is absent in this case.
The reason of that
is an inner structure (i.e., dynamical form-factors) of neutral
composites in these theories. 

In order to be concrete, we will consider
the 
(d + 1)-dimensional 
Nambu-Jona-Lasinio (NJL) models in a strong magnetic
field for arbitrary $d \geq 2$. 
In the regime with the LLL dominance, we derive
the effective action of the corresponding NCFT in the
models with a large number of fermion colors $N$ and analyze
their dynamics. These NCFT are consistent and
quite sophisticated. An especially interesting case is that for
a magnetic field configuration with the maximal number
of independent nonzero tensor components. In that case,
the theories are finite for even $d$ and their dynamics is
quasi-(1+1)-dimensional for
odd $d$ [for even $d$, the NCFT describe a confinement dynamics 
of charged particles].
As will be shown in this paper,
it is the LLL dominance
that provides the exponentially damping (form-)
factors which are 
responsible for finiteness
of these NCFT for even $d$ and their quasi-(1+1)-dimensionality
for odd $d$.
Thus, besides being low
energy theories of the NJL models in a strong magnetic field, the
NCFT based on the LLL dynamics are self-contained and
self-consistent.

We will use two different sets of composite fields for
the description of the dynamics. The first set
uses the conventional 
composite fields $\sigma(x) \sim \bar{\psi}(x) \psi (x)$
and $\pi(x) \sim i\bar{\psi}(x)\gamma_5 \psi (x)$. In this case,
besides the usual
Moyal factor,
additional exponentially damping factors
occur in the interaction vertices of the fields $\sigma(x)$
and $\pi(x)$.
These factors reflect an inner
structure of composites and play an important role in providing
consistency of these NCFT. In particular, because of them,
the UV/IR mixing is absent in these theories. In the second
approach, one
considers other, ``smeared", fields $\Sigma(x)$ and $\Pi(x)$,
connected with $\sigma(x)$ and  $\pi(x)$ through a non-local
transformation. Then, while the additional factors are removed in the
vertices of the smeared fields, they appear in their propagators,
again resulting in the UV/IR mixing removal. By using 
the Weyl symbols of the smeared
fields, we derive the effective action for the composites
in the noncommutative coordinate space.

The paper is organized as follows. In Section \ref{2}, 
in order to understand  the nature of the modified
NCFT in a clear and simple way,
we discuss the quantum mechanical model in a magnetic
field introduced in Ref. \cite{BS}. We show that besides
the solution of Ref. \cite{BS}, which mimics
a conventional NCFT, there is another solution, 
with an interaction vertex containing 
exponentially damping factors.
The existence of these two solutions reflects the possibility of
two different treatments of the case with the particle mass 
$m \to 0$
in this model. In Section \ref{3}, the effective action of
the NCFT connected with the (3+1)-dimensional NJL model in
a strong magnetic field is derived. In Section \ref{4},
the dynamics of this model is discussed. In Section \ref{5},
we generalize the analysis to a general case of $d+1$ dimensions 
with $d \geq 2$. In Section \ref{6}, we summarize the main
results of the paper. In Appendices A and B, some useful
formulas and relations are derived.

\section{Nonrelativistic model}
\label{2}

In order to understand better the nature of the modified NCFT, in this
section we analyze a simple quantum mechanical 
two-dimensional system: 
a pair of unit charges of opposite sign (i.e., a dipole)
in a constant magnetic field and with a harmonic potential interaction 
between them. This model was considered in Ref. \cite{BS}.  
It was argued there that for a strong magnetic field 
this simple 
system reproduces the dynamics
of open strings
attached to D-branes in antisymmetric tensor backgrounds. 

We will show
that important features of the modified NCFT occur already in this simple
quantum mechanical model. Its Lagrangian reads
\ba
L = \frac{m}{2}(\vec{\dot{x}}\,_1^2 + \vec{\dot{x}}\,_2^2) +
\frac{eB}{2}\epsilon^{ab}(\dot{x}_1^ax_1^b - \dot{x}_2^ax_2^b) -
\frac{K}{2}(\vec{x}_1 - \vec{x}_2)^2, \,\,\,\,\,\,\,\,\,\,\,
\epsilon^{ab} = \left(%
\begin{array}{cc}
  0 & 1 \\
  -1 & 0 \\
\end{array}%
\right)\,.
\label{lagrangian1}
\ea
It is convenient to use the center of mass and relative coordinates,
$\vec{X}=\frac{\vec{x}_1 + \vec{x}_2}{2}$
and $\vec{\Delta}=\frac{\vec{x}_1 - \vec{x}_2}{2}$. In these coordinates, 
Lagrangian (\ref{lagrangian1}) takes the form
\ba
L = m(\vec{\dot{X}}^2 + \vec{\dot{\Delta}}^2) +
2eB\epsilon^{ab}\dot{X^a}\Delta^b  -
2K\vec{\Delta}^2.
\lb{lagrangian2}
\ea
The LLL dominance occurs when either $B \to \infty$ or 
$m \to 0$. 
Taking $m = 0$,
the authors of \cite{BS} drop the kinetic terms in 
Lagrangian (\ref{lagrangian2}) 
that results in a theory of the Chern--Simons
type with only first order time derivatives. Then they introduce 
an additional potential $V(\vec{x}_1)$ describing 
an interaction of the first charge with an ``impurity" centered
at the origin  
and show that the matrix element of
$V(\vec{x}_1)$ between dipole states contains the usual Moyal phase
that is a signature of NCFT.

Notice that this result is obtained when the limit $m \to 0$ is
taken directly in the Lagrangian. Let us now show that when one 
first solves 
this
problem for a nonzero $m$ and then takes the limit $m \to 0$ 
in the solution, additional exponential factors occur in the
matrix element of $V(\vec{x_{1}})$.

The Hamiltonian in model (\ref{lagrangian1}) is given by
\ba
H = \frac{\hat{\vec{p}}\,^2 + \hat{\vec{d}}\,^2}{4m} -
\frac{eB}{m}\epsilon^{ab}\Delta^a \hat{p}^b + (\frac{e^2B^2}{m} 
+ 2K)\vec{\Delta}^2,
\label{hamiltonian}
\ea
where $\hat{\vec{p}}$ and $\hat{\vec{d}}$ are operators of the center 
of mass and relative momenta. Since the Hamiltonian 
is independent
of the center of mass coordinates, the wave function can be 
represented in the form $\psi(\vec{X},\vec{\Delta}) 
= e^{i\vec{p}\vec{X}} f(\vec{\Delta})$.
Then, for $f(\vec{\Delta})$ we get the equation
\ba
\left(\frac{\hat{\vec{d}}\,^2}{4m} - \frac{eB}{m}\epsilon^{ab}\Delta^a p^b 
+ (\frac{e^2B^2}{m} + 2K)\vec{\Delta}^2\right) f(\vec{\Delta}) =
(E - \frac{\vec{p}^2}{4m})f(\vec{\Delta}).
\label{dipoles-equation1}
\ea
Changing the variables to
$$
x = \Delta_x + \frac{eBp_y}{2(e^2B^2 + 2Km)},
$$
$$
y = \Delta_y - \frac{eBp_x}{2(e^2B^2 + 2Km)},
$$
we arrive at the equation
\ba
\left(\frac{K\vec{p}\,^2}{2(e^2B^2 + 2Km)} - 
\frac{\partial_x^2 + \partial_y^2}{4m} + (\frac{e^2B^2}{m} + 2K)
(x^2+y^2)\right) f(x,y) = E f(x,y).
\label{dipoles-equation2}
\ea
Clearly, the first term here is the kinetic energy of the center of mass. 
Note that as $m \to 0$, it coincides with the eigenvalue of the 
Hamiltonian
in Ref. \cite{BS} obtained from Lagrangian (\ref{lagrangian1}) with 
$m = 0$. 
All other terms that are present in 
Hamiltonian (\ref{dipoles-equation2}) [and reflecting the inner structure 
of composite states] are absent in the Hamiltonian of Ref. \cite{BS}.
In that case, the only information about the inner structure of 
composites that is retained is given by the relation
\ba
\Delta^a = -\frac{\epsilon^{ab}p^b}{2eB},
\lb{constraint}
\ea
which expresses relative coordinates through the center of mass momentum.

Obviously, equation (\ref{dipoles-equation2}) admits an exact solution. 
Its spectrum contains an infinite number
of composites (neutral bound states) with the energy eigenvalues
\ba
E_{\vec{p},n,k} = \frac{K\vec{p}\,^2}{2r^2} + (n + k + 1) \frac{r}{m},
\lb{spectrum} 
\ea
where $r = \sqrt{e^2B^2+2Km}$ and
$n$ and $k$ are positive integers or zero. Note that in the 
limit $K \to 0$ the Lagrangian (\ref{lagrangian1}) reduces to the 
Lagrangian
of two noninteracting charged particles in a constant magnetic field 
(the Landau problem) and Eq.(\ref{spectrum}) correctly reproduces
the Landau spectrum.

Thus, the model (\ref{lagrangian1}) 
describes an infinite number of neutral composites. 
The vector $\vec{p}$ is their center of 
mass momentum and the last term in (\ref{spectrum}) 
reflects their nontrivial inner structure. 
Now, in the limit $ m \to 0$,
only the LLL states with $n=k=0$ survive 
(all higher excitations decouple). The normalized LLL wave function
with the center of mass momentum $\vec{p}$ is given by
\ba
<\vec{X},\vec{\Delta}|\vec{p}>\, 
=\, \psi_{\vec{p},0,0}(\vec{X},\vec{\Delta}) 
= \left( \frac{r}{2\pi^3} \right)^{1/2} e^{i\vec{p}\vec{X}}
e^{-r(\Delta_x+\frac{eBp_y}{2r^2})^2}e^{-r(\Delta_y-\frac{eBp_x}{2r^2})^2}.
\lb{wavefunctions}
\ea
The Gaussian exponential factors here reflect the inner 
structures of the composites. It is important that
in the limit $m \to 0$, this wave function does {\it not} 
coincide with the wave function of Ref. \cite{BS} corresponding to
the model with the Lagrangian (\ref{lagrangian1}) at $m = 0$:
there are no Gaussian exponential factors in that case. In other
words, while in the $m \to 0$ model, there are quantum
fluctuations described by the Gaussian exponents, these
fluctuations are completely suppressed in the
model with $m \equiv 0$.

Thus we conclude that the quantum dynamics in the limit $m \to 0$ 
in the massive
model does not coincide with that in the massless one. Recall that
the same situation takes place in non-abelian gauge theories: the
limit $m \to 0$ in a massive non-abelian model does not yield
the dynamics of the massless one \cite{gauge}. The origin of this
phenomenon is the same in both cases. Because of constraints
in the massless models, the number of physical degrees of
freedom there is less than the number of degrees of freedom in
the massive ones. In the present quantum mechanical model, these
constraints are described by equation (\ref{constraint}).

Of course, there is nothing wrong with the model
(\ref{lagrangian1}) at $m = 0$. It is
mathematically consistent. However, its dynamics is very
different from that of a physical dipole in a strong
magnetic field [by a physical dipole, we understand
a dipole composed of two massive charged particle,
including the case of an infinitesimally small mass
$m \to 0$]. We also would like to point out that 
the present treatment of the dynamics in a strong magnetic field
is equivalent to the formalism
of the projection onto the LLL developed in Refs. \cite{GK,DJ}.

If following Ref. \cite{BS} we introduce an 
additional potential $V(\vec{x}_1)$ describing an interaction of the
first charge with an
impurity, the matrix element
$<\vec{k}|V(\vec{x}_1)|\vec{p}>$ will describe the scattering 
of composites on the impurity in 
the Born approximation. 
In order to evaluate $<\vec{k}|V(\vec{x}_1)|\vec{p}>$,
it is convenient to introduce the Fourier transform
\ba
V(\vec{x}_1) = \int 
\frac{d^{2}q}{(2\pi)^2} \tilde{V}(\vec{q})e^{i{\vec{q}}{\vec{x}_1}},
\ea
so that
\ba
<\vec{k}|V(\vec{x}_1)|\vec{p}> = \int
\frac{d^{2}q}{(2\pi)^2} \tilde{V}(\vec{q})
<\vec{k}|e^{i\vec{q}(\vec{X}+\vec{\Delta})}|\vec{p}>.
\ea

Inserting now a complete set 
$\int d^2X d^2\Delta |\vec{X}, \vec{\Delta}><\vec{X}, \vec{\Delta}|$
and using Eq. (\ref{wavefunctions}), 
one can easily calculate the matrix element 
$<\vec{k}|e^{i\vec{q}(\vec{X}+\vec{\Delta})}|\vec{p}>$.
In the limit $m \to 0$, it is:
\ba
<\vec{k}|e^{i\vec{q}(\vec{X}+\vec{\Delta})}|\vec{p}> 
= \delta^2(\vec{k} - \vec{q} - \vec{p})
e^{-\frac{\vec{q}^2}{4|eB|}}e^{-\frac{i}{2}q \times k}
\lb{me}
\ea
with the cross product 
$q \times k \equiv \epsilon^{ab}q^{a}k^{b}/eB$.
One can see that, in addition to the standard Moyal factor
$e^{-\frac{i}{2}q \times k}$,
this vertex contains also the exponentially damping term 
$e^{-\frac{\vec{q}^2}{4|eB|}}$. This term of course occurs 
due to the Gaussian factors in the wave function (\ref{wavefunctions}). 
It would be absent if we,
as in Ref. \cite{BS}, used the Lagrangian with
$m = 0$ in Eq. (\ref{lagrangian1}).

The general character of this phenomenon suggests that additional
exponential terms in interaction vertices should also occur in 
field theories 
in a strong magnetic filed. This expectation 
will be confirmed in the next 
Section 
where it will be also shown that these theories
determine a class of modified NCFT.

\section{The NJL model in a magnetic field
as a NCFT: The effective action }
\label{3}

In this Section, we will consider the dynamics in the
(3 + 1)-dimensional
NJL model in a strong magnetic field.
Our aim is to show that this dynamics determines a consistent NCFT.  
As it will be shown in Section \ref{5}, a similar situation takes place 
in an arbitrary 
dimension $D = d + 1$ with the space dimension $d \geq 2$.

The Lagrangian density of the NJL model with the $U_L(1) \times U_R(1)$ 
chiral
symmetry reads
\ba
L = \frac{1}{2}[\bar{\psi}, (i\gamma^{\mu}D_{\mu})\psi]  +
\frac{G}{2N} \left[(\bar{\psi}\psi)^2 + (\bar{\psi}i\gamma^5\psi)^2 
\right],
\label{NJLaction}
\ea
where fermion fields carry an additional ``color" index $i=1,2,...,N$
and the covariant derivative $D_{\mu}$ is
\ba
D_{\mu} = \partial_{\mu} - ieA_{\mu}^{ext}.
\ea
The external vector potential $A_{\mu}^{ext}$ describes a constant
magnetic field $B$ directed in the $+x^3$ direction. We will use
two gauges in this paper: the symmetric gauge
\ba
A_{\mu}^{ext} = (0,\frac{Bx^2}{2},-\frac{Bx^1}{2},0)
\label{symm}
\ea
and the Landau gauge
\ba
A_{\mu}^{ext}  = (0,Bx^2,0,0).
\label{dau}
\ea

We will consider the dynamics of neutral bound states (``dipoles")
in this model with large $N$, when the $1/N$
expansion is justified. In this
case, the model becomes essentially soluble. The central dynamical
phenomenon in the model is the phenomenon of the magnetic catalysis:
a constant magnetic field is a strong catalyst of dynamical chiral
symmetry breaking, leading to the generation of a fermion dynamical mass 
even at the weakest attractive interaction between fermions
\cite{GMS1,GMS2}. The essence of this effect is the dimensional
reduction in the dynamics of fermion pairing in a strong magnetic
field, when the LLL dynamics dominates. In the original
papers \cite{GMS1,GMS2}, the dynamics in the dimensions 
$D = 2 + 1$ and $D = 3 + 1$  were considered. That analysis was
extended to the case of a general space dimension  $d \geq 2$
in Ref. \cite{Gr} [for earlier consideration of dynamical symmetry
breaking in a magnetic field, see Refs. \cite{kw,kl}].  

It is well known that in the model (\ref{NJLaction}) with large
$N$, the relevant neutral degrees of freedom are connected with 
the composite fields $\sigma \sim \bar{\psi}\psi$ and
$\pi \sim \bar{\psi}i\gamma_{5}\psi$.  
The action for them has the following form \cite{GMS2}:
\ba
\Gamma(\sigma, \pi) = -i Tr \mbox{Ln} \left( i\gamma^{\mu}D_{\mu} - 
(\sigma + i\gamma^5\pi) \right) - \frac{N}{2G}\int d^4x(\sigma^2+\pi^2).
\lb{action}
\ea
The gap equation for $<0|\sigma|0>=m$, where $m$ is the fermion dynamical
mass, is
\ba
\frac{\partial}{\partial \sigma}V(\sigma,\pi)|_{\sigma = m, \pi = 0}
= 0,
\ea
where $V(\sigma,\pi)$ is the potential connected with action
(\ref{action}).
According to \cite{GMS2}, in a magnetic field,
this equation has a non-zero solution for
the mass $m$ for an arbitrary positive $G$,
i.e., the critical coupling constant equals zero in this problem. 

The LLL dominance takes place in 
the weak coupling regime, with the dimensionless
coupling constant $g \equiv G\Lambda^2/4\pi^2 \ll 1$. In this case, the
dynamical mass $m$ is \cite{GMS2}
\ba
m^2 = \Lambda^2 e^{-\frac{4\pi^2}{|eB|G}}= 
\Lambda^2 e^{-\frac{\Lambda^2}{|eB|g}},
\lb{solution}
\ea 
where $\Lambda$ is an ultraviolet cutoff connected with 
longitudinal momenta $k_{\|} = (k^0,k^3)$
(we assume that $\Lambda^2 \gg |eB|$). 
\footnote {As will become clear below,
there are no divergences
connected with transverse momenta  $\vec{k}_{\perp} = (k^1,k^2)$
in the regime with the LLL dominance, and therefore 
the ``longitudinal" cutoff removes all divergences in 
this model.}
Notice that
Eq. (\ref{solution}) implies the following hierarchy of scales:
$\frac{|eB|}{m^2} \gg \frac{\Lambda^2}{|eB|}$. 
It will be shown in Section \ref{4} that 
a meaningful continuum limit   
$\Lambda^2 = C|eB| \to \infty$, with $C \gg 1$ and $m$ being fixed,
exists in this model. 

It is straightforward to calculate the interaction vertices 
for the $\tilde{\sigma}=\sigma - m$ and $\pi$ fields
that follows from action (\ref{action}). For example, the 3-point 
vertex $\Gamma_{\tilde{\sigma}\pi\pi}$ is given by
\ba
\Gamma_{\tilde{\sigma}\pi\pi} = 
\int d^4x d^4y d^4z tr[S(x,y)\gamma^5\pi(y)S(y,z)
\gamma^5\pi(z)S(z,x)\tilde{\sigma}(x)].
\label{3point-initial}
\ea
Here the LLL fermion propagator $S(x,y)$ is equal to \cite{GMS2}  
\ba
S(x,y) = e^{\frac{i}{2}(x-y)^{\mu}A_{\mu}^{ext}(x+y)}\tilde{S}(x-y),
\lb{LLLpropagator}
\ea
where the Fourier transform of 
the translationally invariant part $\tilde{S}$ is
\ba
\tilde{S}(k) = i\, e^{-\frac{k_{\perp}^2}{|eB|}}\,
\frac{k^0\gamma^0 - k^3\gamma^3 + m}{k_0^2-k_3^2-m^2}\,
(1-i\gamma^1\gamma^2\mbox{sign}(eB)),
\lb{fourier}
\ea
[for convenience, the $\delta$-symbol with color indices in the 
propagator is omitted]. 
The first factor in (\ref{LLLpropagator}) is 
the Schwinger phase factor \cite{Schwinger}. It breaks the 
translation
invariance even in the case of a constant magnetic 
field, although in this case there is a group of magnetic 
translations whose generators, unlike
usual momenta, do not commute. 

The Schwinger phase 
$\phi = \frac{i}{2}(x-y)^{\mu}A_{\mu}^{ext}(x+y)$ is equal to
\ba
\phi_{sym} = \frac{ieB}{2}\epsilon^{ab}x^ay^b, \,\,\,\,\,\,\,\,\,\,\,
a, b = 1, 2
\label{phase1}
\ea
in the symmetric gauge (\ref{symm}), and it is
\ba
\phi_{Landau} = \phi_{sym} + \frac{ieB}{2}(x^1x^2 - y^1y^2)
\label{phase2}
\ea
in the Landau gauge (\ref{dau}).
One can easily check that the total phase along the closed fermion 
loop in (\ref{3point-initial}) is gauge invariant, i.e., 
independent of a gauge. 

We will show that, in the regime with the LLL dominance,
the effective action (\ref{action}) leads to
a NCFT with noncommutative space transverse coordinates
$\hat{x^a}$:
\ba
[\hat{x^a},\hat{x^b}] = i\frac{1}{eB}\epsilon^{ab} \equiv i\theta^{ab}.  
\label{commrel}
\ea
It is the  
Schwinger phase that is responsible for this noncommutativity. 
Indeed, the commutator 
$[\hat{x^a},\hat{x^b}]$ 
is of course antisymmetric and
the only place where an antisymmetric tensor occurs
in 3-point vertex
(\ref{3point-initial})  
is the Schwinger phase (as will be shown 
below, a similar situation takes place also for higher 
vertices). 

We begin our analysis with the observation that the LLL fermion
propagator (\ref{LLLpropagator}) factorizes into two parts,
the part depending on the transverse coordinates 
$x_{\perp}=(x^1, x^2)$ and that 
depending on the longitudinal coordinates
$x_{\|}=(x^0,x^3)$:  
\ba
S(x,y) = P(x_{\perp},y_{\perp})\, S_{\|}(x_{\|}-y_{\|}).
\lb{factorization}
\ea
Indeed, taking into account expressions (\ref{LLLpropagator}),
(\ref{fourier}), and (\ref{phase1}), we get in the symmetric gauge:
\ba
P(x_{\perp},y_{\perp}) =  \frac{|eB|}{2\pi}\, e^{\frac{ieB}{2}
\epsilon^{ab}x^{a}y^{b}}\,e^{-\frac{|eB|}{4}(\vec{x}_{\perp} 
- \vec{y}_{\perp})^2}
\lb{projector}
\ea
and
\ba
S_{\|}(x_{\|}-y_{\|}) = \int \frac{d^2k_{\|}}{(2\pi)^2}
e^{ik_{\|}(x^{\|}-y^{\|})} \frac{i}{k_{\|}\gamma^{\|} - m}\,
\frac{1 - i\gamma^1\gamma^2 \mbox{sign}(eB)}{2}
\lb{flatspace}
\ea
[henceforth, for concreteness, we will use the symmetric gauge].
The longitudinal part $S_{\|}(x_{\|}-y_{\|})$ is nothing
else but a fermion propagator in 1+1 dimensions. In particular,
the matrix $(1 - i\gamma^1\gamma^2 \mbox{sign}(eB))/2$ is the
projector on the fermion (antifermion) states with the spin
polarized along (opposite to) the magnetic field, and therefore it 
projects on two states of the four ones, as should be in 1+1 
dimensions. As to the operator $P(x_{\perp},y_{\perp})$, it 
is easy to check that it satisfies the relation
\ba
\int d^{2}y^{\perp} P(x^{\perp},y^{\perp})\,P(y^{\perp},z^{\perp}) =
P(x^{\perp},z^{\perp})
\lb{Pprojector}
\ea
and therefore is a projection operator.
Since $S(x,y)$ is
a LLL propagator, it is clear that $P(x_{\perp},y_{\perp})$ is a 
projection operator on the LLL states.

The factorization of the LLL propagator leads to a simple
structure 
of interaction vertices for $\pi$ and
$\tilde{\sigma}$ fields. For example, as it
is clear from expression (\ref{3point-initial}) 
for the 3-point vertex, a substitution of the 
Fourier transforms
for the fields makes the integration over the
longitudinal and transverse coordinates
completely independent. And since $S_{\|}$ is a
(1 + 1)-dimensional propagator, the integration over $x_{\|}$ 
coordinates yields
a fermion loop in the (1+1)-dimensional Minkowski space.
It is obvious that the same is true also for higher order interaction
vertices arising from action (\ref{action}). Therefore the dependence
of interaction vertices on longitudinal 
$k_{\|}$ momenta is standard and, for clarity of
the presentation, we will first
consider the case with all external longitudinal
momenta entering the fermion loop to be
zero. This of course corresponds to the choice of
$\pi$ and $\tilde{\sigma}$ fields independent of  
longitudinal coordinates $x_{\|}$. The general case,
with the fields depending on both transverse and longitudinal
coordinates, will be considered in the end of this Section.

Now, substituting 
the Fourier transforms of the fields $\pi(x_{\perp})$ and 
$\tilde{\sigma}(x_{\perp})$ into Eq. (\ref{3point-initial}) 
and using Eqs. (\ref{factorization}), (\ref{projector}), and
(\ref{flatspace}), 
we find the following expression for the 3-point interaction
vertex $\Gamma_{\tilde{\sigma}\pi\pi}$ in the momentum space:
$$
\Gamma_{\tilde{\sigma}\pi\pi} = -\frac{N|eB|}{m} \int d^2x_{||}
\int \frac{d^2k_1d^2k_2d^2k_3}{(2\pi)^6} 
\pi(k_1)\pi(k_2)\tilde{\sigma}(k_3)\,
\delta^2(k_1+k_2+k_3)
$$
\ba
\times \,\,\, e^{-\frac{k_{1}^2+k_{2}^2+k_{3}^2}{4|eB|}}\,
\mbox{exp}[-\frac{i}{2}(k_1 \times k_2 + k_1 \times k_3 + 
k_2 \times k_3)],
\lb{3point}
\ea
where $k_i \times k_j = k_{i}^a \theta^{ab}k_{j}^b 
\equiv k_{i} \theta k_{j}$, 
$\theta^{ab}=  \frac{1}{eB}\epsilon^{ab}$ (here, for convenience,
we omitted the subscript $\perp$ for the transverse coordinates).
Notice that because of the exponentially damping factors, there
are no ultraviolet divergences in this expression. 

According to \cite{DNS}, an $n$-point vertex in a 
noncommutative theory in momentum space has the following
structure:
\ba
\int \frac{d^Dk_1}{(2\pi)^D} ... \frac{d^Dk_n}{(2\pi)^D} 
\phi(k_1)... \phi(k_n) \delta^D(\sum_i k_i)
e^{-\frac{i}{2} \sum_{i<j} k_i \times k_j},
\label{vertex}
\ea
where here $\phi$ denotes a generic field and the exponent 
$e^{-\frac{i}{2} \sum_{i<j} k_i \times k_j}\equiv
e^{-\frac{i}{2} \sum_{i<j} k_i \theta k_j}$ 
is the Moyal exponent
factor. Comparing expressions (\ref{3point}) and (\ref{vertex}),
we see that apart from the factor  
$e^{-\frac{k_{1}^2+ k_{2}^2+k_{3}^2}{4|eB|}}$,
the vertex $\Gamma_{\tilde{\sigma}\pi\pi}$ coincides with the standard 
3-point
interaction vertex in a noncommutative theory with the commutator 
$[\hat{x^a},\hat{x^b}]= i\theta^{ab} = \frac{i}{eB}\epsilon^{ab}$.
In order to take properly into account this additional factor in 
the vertex,
it will be convenient to introduce new, ``smeared'', fields:
\ba
\Pi(x) = e^{\frac{\nabla_{\perp}^2}{4|eB|}}\,\pi(x),\,\,
\Sigma(x) = e^{\frac{\nabla_{\perp}^2}{4|eB|}}\,\sigma(x),
\label{smeared}
\ea
where $\nabla_{\perp}^2$ is the transverse Laplacian.
Then, in terms of these fields, the vertex can be rewritten in the
standard form with the Moyal exponent factor:
\ba
\Gamma_{\tilde{\Sigma}\Pi\Pi} = -\frac{N|eB|}{m} \int d^2x_{||}
\int \frac{d^2k_1d^2k_2d^2k_3}{(2\pi)^6}
\Pi(k_1)\Pi(k_2)\tilde{\Sigma}(k_3)\,\delta^2(\sum_{i} k_i)\,
\exp[-\frac{i}{2}\sum_{i<j} k_i \times k_j].
\lb{3point1}
\ea
One can similarly analyze
the 4-point interaction vertex $\Gamma_{4\Pi}$. We get:
\ba
\Gamma_{4\Pi} = -\frac{N|eB|}{4m^2} \int d^2x_{||} \int
\frac{d^2k_1 d^2k_2 d^2k_3 d^2k_4}{(2\pi)^8} \Pi(k_1)\Pi(k_2)
\Pi(k_3)\Pi(k_4) \delta^2(\sum_{i} k_i)\, \exp[-\frac{i}{2}
\sum_{i<j} k_i \times k_j].
\label{4point}
\ea
The occurrence of the smeared 
fields in the vertices reflects
an inner structure (dynamical form-factors)
of $\pi$ and $\sigma$ composites on the lowest
Landau level, which is
similar to that of a dipole in the quantum mechanical problem
considered in Section \ref{2}.

As is well known, the cross product in the momentum 
space corresponds to a star product in the coordinate space
\cite{DNS}:
\ba
(\Phi * \Phi)(x) =
e^{\frac{i}{2}\theta^{ab}
\frac{\partial}{\partial y^a}\frac{\partial}{\partial z^b}}
\Phi(y)\Phi(z)|_{y=z=x},
\lb{starproduct}
\ea
where here $\Phi$ represents the smeared fields $\Pi$ and 
$\Sigma$.
By using the star product, one can rewrite the vertices
$\Gamma_{\tilde{\Sigma}\Pi\Pi}$ and $\Gamma_{4\Pi}$
in the following simple form in the coordinate space:
$$
\Gamma_{\tilde{\Sigma}\Pi\Pi} = -\frac{N|eB|}{4\pi^2m}
\int d^2x_{||}d^2x_{\perp}\,\,\,
\tilde{\Sigma}*\Pi*\Pi\,,
$$
\ba
\Gamma_{4\Pi} = -\frac{N|eB|}{16\pi^2m^2} \int
d^2x_{||}d^2x_{\perp}\,\,\, \Pi*\Pi*\Pi*\Pi.
\label{xspacevertices}
\ea

As to expressing the vertices in NCFT in
the space with noncommutative coordinates $\hat{x^a}$, 
one should use the Weyl symbol of a field $\Phi$ there
\cite{DNS}:
\ba
\hat{\Phi}(\hat{x}) \equiv \hat{W}[\Phi] =
\int d^Dx\,\Phi(x)\,\hat{\Delta}(x),\,\,\,\,\,\,\,\,\,
\hat{\Delta}(x) \equiv \int \frac{d^Dk}{(2\pi)^D} e^{ik_a\hat{x}^a}\,
e^{-ik_ax^a}.
\label{Weylsymbol}
\ea
The most 
important
property of the Weyl symbol is that the product of the Weyl symbols of 
two 
functions is equal to the Weyl symbol of their star product:
\ba
\hat{W}[\Phi_1]\,\hat{W}[\Phi_2] = \hat{W}[\Phi_{1}*\Phi_{2}].
\label{W1}
\ea
In our case, the Weyl symbol $\hat{\Phi}$ represents  
$\hat{\Pi}$ and $\hat{\tilde{\Sigma}}$. 
Note that the relation between the Weyl symbols of
smeared and non-smeared fields is:
\ba
\hat{\Phi}(\hat{x}) =
e^{\frac{\hat{\nabla}_{\perp}^2}{4|eB|}}\,\hat{\phi}(\hat{x}),
\label{Wsmeared}
\ea
where the operator $\hat{\nabla}_{\perp}^2$ in the noncommutative
space acts as
\ba
\hat{\nabla}_{\perp}^2\,\hat{\phi}(\hat{x}) =
-(eB)^2\,\sum_{a = 1}^2
[\hat{x}^a,[\hat{x}^a, \hat{\phi}(\hat{x})]\,]
\ea
[the latter relation follows from the definition of the derivative
in NCFT, 
$ \hat{\nabla}_{\perp a}\,\hat{\phi}(\hat{x})=
-i[(\theta^{-1})_{ab}\hat{x}^b, \hat{\phi}(\hat{x})]$ \cite{DNS}\,].

In terms of $\hat{\Phi}$, the 3- and 4-point
vertices take the following form in NCFT:
$$
\Gamma_{\tilde{\Sigma}\Pi\Pi} = -\frac{N|eB|}{4\pi^2m}\,
\int d^2x_{||}\mbox{\bf Tr}\,
\hat{\tilde{\Sigma}}\hat{\Pi}^2\,,
$$
\ba
\Gamma_{4\Pi} = -\frac{N|eB|}{16\pi^2m^2}\,
\int d^2x_{||}\mbox{\bf Tr}\,\hat{\Pi}^4\,,
\label{ncspacevertices}
\ea
where the operation $\mbox{\bf Tr}$ is defined as in \cite{DNS}.
As is shown in Appendix A, all interaction
vertices $\Gamma_{n\Phi}$ ($n \geq 3$) arising from action
(\ref{action})
have a similar structure. 

There exists another, more convenient for practical calculations,
representation of interaction vertices in which the vertices
are expressed through the initial, ``non-smeared", fields
$\pi$ and $\tilde{\sigma}$.
The point is that,
due to the presence of the $\delta$-function 
$\delta^2(\sum_i k_i)$, 
the exponent factors
$e^{-\frac{\sum_{i=1}^n\vec{k}_i^2}{4|eB|}}
e^{\frac{-i}{2}\sum_{i < j} {k_i \times k_j}}$ in an
n-point vertex can be rewritten as 
$e^{-\frac{i}{2} \sum_{i<j} k_i \times_M k_j}$, where
$k_i \times_M k_j$ is a new cross product. It is defined as
\ba
k_i \times_M k_j = k_i \Omega k_j
\lb{Mcross}
\ea
with the matrix $\Omega$ being
\ba
\Omega^{ab} = \frac{1}{|eB|} \left(\begin{array}{c}
i\,\,\,\,\,\,\,\,\,\mbox{sign}(eB) \\
-\mbox{sign}(eB)\,\,\,\,\,\,\,i \end{array}\right).
\ea
We will call $k_i \times_M k_j$ an $M$ (magnetic)-cross product.
Notice that like
the matrix $\theta^{ab}$, defining the cross product,
the new matrix
$\Omega^{ab}$, defining the $M$-cross product, is
anti-hermitian.

By using the $M$-cross product, we get the following simple structure
for an n-point vertex in the momentum space: 
\ba
\int d^2x_{||}\frac{d^2k_1}{(2\pi)^2} ... \frac{d^2k_n}{(2\pi)^2}
\phi(k_1)... \phi(k_n) \delta^2(\sum_i k_i)
e^{-\frac{i}{2} \sum_{i<j} k_i \times_{M} k_j}
\label{vertex1}
\ea
(compare with expression (\ref{vertex})). Here the 
field $\phi$ represents initial fields $\pi$ and $\tilde{\sigma}$. 

In the coordinate space, the $M$-cross product becomes
an $M$-star product:
\ba
(\phi *_M \phi)(x) =
e^{\frac{i}{2}\Omega^{ab}
\frac{\partial}{\partial y^a}\frac{\partial}{\partial z^b}}
\phi(y)\phi(z)|_{y=z=x}.
\lb{Mproduct}
\ea
(compare with equation (\ref{starproduct})).
By using the $M$-star product, one can express 
n-point vertices  
through the initial $\pi$ and $\tilde{\sigma}$ fields
in the coordinate space. For example, the vertices 
$\Gamma_{\tilde{\sigma}\pi\pi}$ and $\Gamma_{4\pi}$
become:
$$
\Gamma_{\tilde{\sigma}\pi\pi} = -\frac{N|eB|}{4\pi^2m} 
\int d^2x_{||}d^2x_{\perp}\,\,\,
\tilde{\sigma}*_M\pi*_M\pi\,,
$$
\ba
\Gamma_{4\pi} = -\frac{N|eB|}{16\pi^2m^2} \int 
d^2x_{||}d^2x_{\perp}\,\,\, \pi*_M\pi*_M
\pi*_M\pi.
\label{xspacevertices1}
\ea

In fact, by using the $M$-star product, the whole 
effective action (\ref{action}) 
can be written in a compact and explicit form  
for the case of fields independent of longitudinal coordinates
$x_{||}$. First, note that for constant fields, the $M$-star product
in $\Gamma_{n\phi}$ vertices 
(\ref{xspacevertices1}) is reduced to
the usual product and the
vertices come from the effective potential in that case. Then, 
this  
implies that, up to the measure $-\int d^4x$,
the whole effective action for fields depending on transverse coordinates 
coincides with the
effective potential 
in which the usual product
is replaced by the $M$-star product in the part coming from the
$Tr \mbox {Ln}$ term in (\ref{action}). 
As to the last
term $\frac{N}{2G} \int d^4x\, (\sigma^2 + \pi^2)$ there, 
it should
stay as it is. This is because
unlike the star product, the $M$-star product
and the usual one lead to different quadratic terms in the action.
Now, by using expression
(\ref{LLLpropagator}) for the fermion propagator, we easily find
the effective potential:
\ba
V(\sigma,\pi) = \frac{N|eB|}{8\pi^2} 
[\sigma^2 + \pi^2]\,[\ln(\frac{\sigma^2 + \pi^2}
{\Lambda^2}) - 1] 
+\frac{N}{2G}(\sigma^2 + \pi^2) + O(\frac{\sigma^2 + \pi^2}{\Lambda^2}).
\label{effpotential}
\ea
Then, the
effective action reads:
\ba
\Gamma(\sigma,\pi) = - \frac{N|eB|}{8\pi^2} \int d^4x
\left([\sigma^2 + \pi^2]\,[\ln(\frac{\sigma^2 + \pi^2}
{\Lambda^2}) - 1] \right)_{*_M}
- \frac{N}{2G} \int d^4x\, \left(\sigma^2
+ \pi^2 \right).
\label{Maction}
\ea
This expression is very convenient for calculating the n-point
vertices $\Gamma_{n\phi}$. In Appendix B, it is shown that
the $M$-star product also naturally appears in the formalism of the
projected density operators developed in Ref. \cite{Sinova}
for the description of the quantum Hall effect.

While the $M$-star product is useful for practical calculations,
its connection with the multiplication operation in 
a noncommutative coordinate space is not direct. This is in
contrast with the star product for which relation (\ref{W1}) 
takes place. Therefore it will be useful to rewrite
action (\ref{Maction}) through the star product. It can be
done by using the smeared fields $\Sigma$ and $\Pi$. The
result is:
\ba
\Gamma = - \frac{N|eB|}{8\pi^2} \int d^4x
\left([\Sigma^2 + \Pi^2]\,[\ln(\frac{\Sigma^2 + \Pi^2}
{\Lambda^2}) - 1] + \frac{4\pi^2}{G|eB|}(\sigma^2 + 
\pi^2)\right)_{*},
\label{staraction}
\ea
where we used the fact that the star product and
the usual one lead to the same quadratic terms in the action.
Notice that the fields $\sigma$ and $\pi$ are connected 
with the smeared fields through the non-local relation
(\ref{smeared}). This relation implies that
exponentially damping form-factors are built in the propagators of the
smeared 
fields. As a result, their propagators    
decrease rapidly, as 
$\mbox{exp}(-k_{\perp}^2/2|eB|)$, with
$k_{\perp}^2 \to \infty$. As will be shown in the next Section,
this property is in particular responsible for removing the UV/IR
mixing in the model.

By using relation (\ref{W1}),
it is straightforward to rewrite the action in the
NCFT through Weyl symbols:
\ba
\Gamma = - \frac{N|eB|}{8\pi^2} \int d^2x_{||}\mbox{\bf Tr}
\left([\hat{\Sigma}^2 + \hat{\Pi}^2]\,
[\ln(\frac{\hat{\Sigma}^2 + \hat{\Pi}^2}
{\Lambda^2}) - 1] + \frac{4\pi^2}{G|eB|}(\hat{\sigma}^2 + 
\hat{\pi}^2)\right).
\label{ncaction}
\ea
The Weyl symbols $\hat{\sigma}$ and $\hat{\pi}$ are connected
with the Weyl symbols $\hat{\Sigma}$ and $\hat{\Pi}$ through
relation (\ref{Wsmeared}).

Let us now generalize expressions (\ref{Maction}),
(\ref{staraction}), and (\ref{ncaction})
to the case when
fields $\pi$ and $\sigma$ depend on both transverse and
longitudinal coordinates. First, as it
follows from the expansion of the first term 
in action (\ref{action}) in a series in $\pi$ and $\tilde{\sigma}$,
the general $n$-point vertex is given by
$$
\Gamma_{n\phi} = -\frac{(-i)^{n+1}N}{n} \int 
d^2x^{\perp}_1...d^2x^{\perp}_n\,
d^2x^{||}_1...d^2x^{||}_n P(x^{\perp}_1,x^{\perp}_2)...
P(x^{\perp}_n,x^{\perp}_1)\,
$$
\ba
\times \mbox{tr}
\left[S_{||}(x_1-x_2)(\tilde{\sigma}(x_2) +i\gamma^5\pi(x_2))...
S_{||}(x_n-x_1)(\tilde{\sigma}(x_1)+i\gamma^5\pi(x_1))\right]
\label{n-order1}
\ea
[notice that here
the longitudinal part of the fermion propagator $S_{||}(x)$
does not contain color indices].
As was shown above and in Appendix A, the transverse part of 
vertex (\ref{n-order1}) can 
be
expressed through the $M$-star product. Therefore the $n$-point
vertex is 
$$
\Gamma_{n\phi} = -\frac{(-i)^{n+1}N|eB|}{2\pi n} \int d^2x^{\perp}\,
d^2x^{||}_1...d^2x^{||}_n\, 
$$
\ba
\times\,\,\mbox{tr}
\left[S_{||}(x_1-x_2)
(\tilde{\sigma}(x^{\perp},x^{||}_2)+i\gamma^5\pi(x^{\perp},x^{||}_2))...
S_{||}(x_n-x_1)(\tilde{\sigma}(x^{\perp},x^{||}_1)
+i\gamma^5\pi(x^{\perp},x^{||}_1))\right]_{*M}.
\label{n-order2}
\ea
This relation implies that the full effective effective action 
can be written through the $M$-star product as:
\ba
\Gamma(\sigma, \pi) = -\frac{iN|eB|}{2\pi} \int d^2x_{\perp}\,
Tr_{||}\left[{\cal P}\,
\mbox{Ln} \left( i\gamma^{||}\partial_{||} -
(\sigma + i\gamma^5\pi) \right)\right]_{*_M} -
\frac{N}{2G}\int d^4x ({\sigma}^2 + {\pi}^2),
\lb{MACTION}
\ea
where the projector ${\cal P}$ is
\ba
{\cal P} \equiv \frac{1 - i\gamma^1\gamma^2 \mbox{sign}(eB)}{2} 
\lb{P_}
\ea
(compare with action (\ref{Maction})). Here the trace $Tr_{||}$,
related to the longitudinal subspace, is taken in the functional sense. 

As to the form of the
effective action written through the star product and its
form in the noncommutative coordinate space, they are:
\ba
\Gamma = \frac{N|eB|}{2\pi} \int d^2x_{\perp}\,
\left[-iTr_{||}\left[{\cal P}\,
\mbox{Ln} \left( i\gamma^{||}\partial_{||} -
(\Sigma + i\gamma^5\Pi) \right)\right] -
\frac{\pi}{G|eB|}\int d^2x_{||}({\sigma}^2+{\pi}^2)\right]_{*}
\lb{starACTION}
\ea
and
\ba
\Gamma = \frac{N|eB|}{2\pi} \mbox{\bf Tr}\,
\left[-iTr_{||}\left[{\cal P}\,
\mbox{Ln} \left( i\gamma^{||}\partial_{||} -
(\hat{\Sigma} + i\gamma^5\hat{\Pi}) \right)\right] -
\frac{\pi}{G|eB|}\int d^2x_{||}(\hat{\sigma}^2+\hat{\pi}^2)
\right]
\lb{ncACTION}
\ea
(compare with Eqs. (\ref{staraction}) and
(\ref{ncaction}), respectively).

This concludes the derivation of the action of the
noncommutative field theory corresponding to the NJL model
in a strong magnetic field.
In the next Section, we will consider the dynamics 
in this model in more detail.  

\section{The NJL model in a magnetic field
as a NCFT: The Dynamics}
\label{4}

In the regime with the LLL dominance, the dynamics of neutral composites 
is described by quite sophisticated NCFT (\ref{ncACTION}).
In this 
Section, we will
show that in this model $i$) there exists a well defined commutative limit
$|eB| \to \infty$ when $[\hat{x}^a, \hat{x}^b] = 0$;
$ii$) the universality class of the low energy 
dynamics,
with $k_{\perp} \ll \sqrt{|eB|}$, is intimately connected with 
the dynamics in the (1+1)-dimensional Gross-Neveu (GN) model
\cite{GN}; and $iii$) there is no UV/IR mixing. 

The key point in the derivation of action (\ref{ncACTION}) was
the fact that the LLL fermion
propagator (\ref{LLLpropagator}) factorizes into two parts
(see Eq. (\ref{factorization})) and that
its transverse part $P(x_{\perp},y_{\perp})$
is a projection operator on
the LLL states. It is quite remarkable that it exactly coincides
with the projection operator on the LLL states
in nonrelativistic dynamics
introduced for the description of the quantum Hall effect in
Refs. \cite{GK,BES}. 
\footnote {This point is intimately connected with that
the wave functions of the LLL states in the Landau problem are
independent of the fermion mass $m$ (see equation 
(\ref{wavefunctions}) for $K = 0$).} 
Therefore the transverse dynamics in this problem is universal and 
peculiarities of the 
relativistic
dynamics reflect themselves only in the (1 + 1)-dimensional 
longitudinal space.

In order to study the low energy dynamics with $k_{\perp} \ll 
\sqrt{|eB|}$,
it will be instructive 
to consider, as in Ref. \cite{Elias},
the following continuum limit:
$\Lambda^2 = C|eB| \to \infty$, with $C \gg 1$ and $m$ being fixed.
Let us 
consider $n$-point
vertex (\ref{n-order1}) in this limit. Since the projection
operator $P(x^{\perp}_i,x^{\perp}_{i + 1})$ is 
\ba
P(x^{\perp}_{i},x^{\perp}_{i + 1}) =  \frac{|eB|}{2\pi}\, 
e^{\frac{ieB}{2}
\epsilon^{ab}x^{a}_{i}x^{b}_{i+1}}\,e^{-\frac{|eB|}{4}
(\vec{x}^{\perp}_{i}
- \vec{x}^{\perp}_{i+1})^2}
\lb{projector1}
\ea
(see Eq. (\ref{projector})), the point with coordinates  
$x^{\perp}_i = x^{\perp}_{i + 1}$ , $i= 1,..., n-1$, is both a saddle and 
stationary
point in the multiple integral (\ref{n-order1}) in the limit
$|eB| \to \infty$. Therefore, in order to get 
the leading term of the asymptotic 
expansion of that integral, one can put 
$x^{\perp}_i = x^{\perp}_{n} \equiv x^{\perp}$
in the arguments of all the fields $\tilde{\sigma}(x^{\perp}_i) 
+i\gamma^5\pi(x^{\perp} _i)$ there.
Then, by using relation (\ref{Pprojector})
and the equality $P(x^{\perp},x^{\perp})= |eB|/2\pi$, 
we easily
integrate over transverse coordinates in (\ref{n-order1})
and obtain the following asymptotic expression:
$$
\Gamma^{(as)}_{n\phi} = -\frac{(-i)^{n+1}}{n}\frac{N|eB|}{2\pi} \int 
d^2x^{\perp}\,
d^2x^{||}_1...d^2x^{||}_n
$$
\ba
\times \mbox{tr}
\left[S_{||}(x_1-x_2)[\tilde{\sigma}(x^{\perp},x^{||}_2)
+i\gamma^5\pi(x^{\perp},x^{||}_2)]...
S_{||}(x_n-x_1)[\tilde{\sigma}(x^{\perp},x^{||}_1) 
+i\gamma^5\pi(x^{\perp},x^{||}_1)]\right].
\label{asvertex}
\ea
This equation implies that as $|eB| \to \infty$, the leading asymptotic
term in the action is:
\ba
\Gamma^{(as)}(\sigma, \pi) = \frac{|eB|}{2\pi} \int 
d^2x^{\perp}\,
\left[-iNTr_{||}\left[{\cal P}\,
\mbox{Ln} \left( i\gamma^{||}\partial_{||} -
(\sigma + i\gamma^5\pi) \right)\right] -
\frac{N\pi}{G|eB|}\int d^2x_{||}(\sigma^2+\pi^2)\right].
\lb{asympaction}
\ea
This action corresponds to a {\it commutative} field theory,
as should be in the limit $|eB| \to \infty$ [indeed, the
commutator
$[\hat{x^a},\hat{x^b}] = i\frac{1}{eB}\epsilon^{ab}$ goes to
zero as $|eB| \to \infty$]. Also, since there is no hopping
term for the transverse coordinates $x^{\perp}$ in this action,
they just play the role of a label of the fields.  

Let us now compare this action with 
the action of the (1+1)-dimensional GN
model \cite{GN}:
\ba
\Gamma_{GN}(\sigma, \pi) = 
-iNTr
\mbox{Ln} \left( i\gamma^{\mu}\partial_{\mu} -
(\sigma + i\gamma^5\pi) \right) -
\frac{N}{2\tilde{G}}\int d^2x(\sigma^2+\pi^2),\,\,\mu = 0,1,
\lb{GN}
\ea
where $\tilde{G}$ is a dimensionless coupling constant. One can see
that, up to the factor $|eB|/2\pi \int d^2x_{\perp}$, these
two actions coincide if the constant $G$ in (\ref{asympaction})
is identified with $2\pi \tilde{G}/|eB|$. In particular,
with this identification, expression (\ref{solution}) for the
dynamical mass coincides with the expression for $m$
in the GN model, $m^2 = \Lambda^2 e^{-\frac{2\pi}{\tilde{G}}}$.
Also, using Eq. (\ref{solution}), one can
express the coupling constant $G$ in the effective potential
(\ref{effpotential}) through the dynamical mass $m$ and cutoff
$\Lambda$. Then, up to
$O((\sigma^2 + \pi^2)/\Lambda^2)$ terms, we get
the expression independent of the cutoff:
\ba
V(\sigma,\pi) = \frac{N|eB|}{8\pi^2}
[\sigma^2 + \pi^2]\,[\ln(\frac{\sigma^2 + \pi^2}
{m^2}) - 1].
\label{renpotential}
\ea
This renormalized form of the potential coincides with the GN potential. 

As to the factor $|eB|/2\pi \int d^2x_{\perp}$,
its meaning is very simple. Since 
density of the LLL states is equal to $|eB|/2\pi$ , this factor
yields the number of the Landau states on the transverse plane. In
other words, as $|eB| \to \infty$, the model is reduced
to a continuum set of independent   
(1+1)--dimensional GN models, labeled by the
coordinates in the plane perpendicular to the magnetic field.
The conjecture about such a structure of the NJL model
in the limit $|eB| \to \infty$ was made in Ref. \cite{Elias}
and was based on a study of the effective potential and
the kinetic (two derivative) term in the model. The present approach
allows to derive the whole action and thus to prove the conjecture. 

The existence of the physically meaningful limit $|eB| \to \infty$
is quite noticeable. It confirms that the model with
the LLL dominance is self-consistent.
In order to understand its dynamics better, it 
is instructive to look at the dispersion relations for
$\sigma$ and $\pi$ excitations with momenta
$k_{\perp} \ll \sqrt{|eB|}$ \cite{GMS2}:
$$
E_\pi \simeq \biggl[\frac{m^2}{|eB|}
\ln\biggl(\frac{|eB|}{\pi m^2}\biggr)
{\vec k}_{\perp}^2+k_3^2\biggr]^{1/2},
$$
\ba
E_{\sigma} \simeq \biggl[12~m^2 +
\frac{3m^2}{|eB|}\ln\biggl(\frac{|eB|}{\pi m^2}\biggr)
{\vec k}^2_{\perp}+k_3^2 \biggr]^{1/2}.
\lb{dispersion}
\ea
We find from these relations that the transverse
velocity $|{\vec v}_{\perp}|=|\partial 
E_{\pi,\sigma}/\partial {\vec k}_{\perp}|$
of both $\pi$ and $\sigma$
goes rapidly (as $O(m^2/|eB|$)) to zero as
$|eB| \to \infty$. In other words, there is no hopping 
between different transverse points in this limit. 
For a strong but finite magnetic field,
the transverse velocity is, although nonzero, very small.
In this case,
the $\pi$ and $\sigma$ composites have a string-like shape:
while their transverse size is of the order of the magnetic
length $l = 1/\sqrt{|eB|}$, the longitudinal size is of order
$1/m$, and $l \ll 1/m$. 

The important point is that besides being a low energy theory
of the initial NJL model in a magnetic field, this truncated
[based on the LLL dynamics] model is self-contained. In particular, in
this model
one can consider arbitrary large values for transverse momenta
$k_{\perp}$, although in this case its dynamics is very
different from that of the initial NJL model. In fact, by
using the expression for the pion propagator (\ref{propagator})   
written below,
it is not
difficult to check that for $k_{\perp} \gg \sqrt{|eB|}$
the dispersion relation for $\pi$ excitations takes the
following form:
\ba
E_{\pi} \simeq \left[4m^2(1-\frac{\pi^2e^{-
\frac{\vec{k}_{\perp}^2}{|eB|}}}
{\ln^2\frac{|eB|}{m^2}}) + k_3^2 \right]^{1/2}.
\lb{dispersion1}
\ea
In this regime, the transverse velocity $|\vec{v}_{\perp}|$ 
is extremely small,
$|\vec{v}_{\perp}| \sim \frac{m|\vec{k}_{\perp}|}{|eB|}
e^{-\vec{k}_{\perp}^2/|eB|}$, and a $\pi$ excitation
is a loosely bound state
moving along the $x_3$ direction. Its mass is close to the $2m$
threshold.

Thus we conclude that the NJL model in a strong magnetic field 
yields an example of a consistent NCFT
with quite nontrivial dynamics. The point that 
exponentially damping factors 
occur either in vertices (for the fields $\sigma$ and $\pi$) or
in propagators (for the smeared fields) plays a crucial role in its 
consistency. Let us 
now
show that these factors are in particular responsible
for removing a UV/IR mixing, the phenomenon that plagues
conventional nonsupersymmetric NCFT \cite{MRS}.

The simplest example of the UV/IR mixing is given by a one-loop 
contribution in a propagator in the noncommutative $\phi^4$ model 
with the action 
\ba
S = \int d^4x \left(\frac{1}{2}(\partial_{\mu}\phi)^2 - 
\frac{m^2\phi^2}{2} - \frac{g^2}{4!}\phi * \phi * \phi * \phi\right).
\ea
There are planar and nonplanar one-loop 
contributions in the propagator of $\phi$ in this model \cite{MRS}:
\ba
\Gamma^{(2)}_{nc} = \Gamma^{(2)}_{pl} + \Gamma^{(2)}_{npl} 
= \frac{g^2}{3(2\pi)^4} \int \frac{d^4k}{k^2+m^2} +  
\frac{g^2}{6(2\pi)^4} \int \frac{d^4k}{k^2+m^2} e^{ik \times p}.
\label{one-loop}
\ea
The nonplanar contribution is specific 
for a noncommutative theory and is responsible for the UV/IR mixing. 
Indeed, the nonplanar contribution is equal to 
\ba
\Gamma^{(2)}_{npl} = \frac{g^2}{96\pi^2}(\Lambda_{eff}^2 
- m^2\ln(\frac{\Lambda_{eff}^2}{m^2}) + O(1)),
\label{npl}
\ea
where
$$
\Lambda_{eff}^2  = \frac{1}{1/\Lambda^2 - p^{i}\theta_{ij}^2p^{j}}
$$
with $\Lambda$ being 
cutoff. It is clear that if the external 
momentum $p \to 0$, the nonplanar contribution  
(\ref{npl}) diverges quadratically. On the other hand,
for a nonzero $p$, it is finite due 
to the Moyal phase factor $ e^{ik \times p}$ in
the second term in expression (\ref{one-loop}) (which oscillates rapidly
at large $k$).
Thus, although the Moyal
factor regularizes the UV divergence, it 
leads to an IR divergence of the integral, i.e., to 
the UV/IR mixing.

Let us now show how the exponentially damping factors in
vertices (for the fields $\sigma$ and $\pi$)
or in propagators (for the smeared fields $\Sigma$ and $\Pi$) 
remove the UV/IR mixing. 
We will first consider the description using the fields $\sigma$
and $\pi$.
As an example,
we will consider the one-loop correction in the $\pi$ propagator 
generated by the four-point interaction vertex $\Gamma_{4\pi}$. 
First, from action (\ref{action}), we get this propagator  
in tree approximation. In Euclidean space it is: 
\ba
D_{\pi}^{(tree)}(p) \simeq 
\frac{4\pi^2}{N|eB|[(1-e^{-\frac{p_{\perp}^2}{2|eB|}})
\ln \frac{|eB|}{m^2} + 
e^{-\frac{p_{\perp}^2}{2|eB|}} \int_0^1 du \frac{p_{\|}^2 u}{p_{\|}^2 
u(1-u) 
+ m^2}]}.
\label{propagator}
\ea
Then, by using this $D_{\pi}^{(tree)}(p)$
and Eq. (\ref{n-order2}) for the 
vertex $\Gamma_{4\pi}$,
we find the following one-loop nonplanar contribution to 
the propagator:
\ba
\frac{N|eB|}{4\pi^3} \int \frac{d^4 k}{(2\pi)^4} e^{-\frac{p_{\perp}^2+
k_{\perp}^2}{2|eB|}} e^{\frac{i}{eB}(p^1k^2-p^2k^1)} 
I(p_{\|},k_{\|}) D_{\pi}^{(tree)}(k),
\label{nonplanar1}
\ea
where
\ba
I(p_{\|},k_{\|}) = \int d^2l_{\|} \frac{(l_{\|}^2+m^2+l_{\|}\cdot p_{\|})
[(p_{\|}-k_{\|}+l_{\|})^2+m^2-(p_{\|}-k_{\|}+l_{\|})\cdot p_{\|}]   
+p_{\|}^2m^2}{(l_{\|}^2+m^2)[(p_{\|}+l_{\|})^2+m^2]
[(p_{\|}-k_{\|}+l_{\|})^2+m^2][(l_{\|}-k_{\|})^2+m^2)]}\,.
\nonumber
\ea
Here the integral over transverse momenta $k_{\perp}$ is
\ba
\int \frac{d^2 k_{\perp}}{(2\pi)^2} 
e^{-\frac{p_{\perp}^2+k_{\perp}^2}{2|eB|}} 
e^{\frac{i}{eB}(p^1k^2-p^2k^1)}D_{\pi}^{(tree)}(k).
\label{integr}
\ea
It is clear that due to the presence of the 
factor $e^{-k_{\perp}^2/2|eB|}$ and because $D_{\pi}^{(tree)}(k)$ is
finite as $k_{\perp}^2 \to \infty$, 
this integral is convergent
for all values of $p_{\perp}$, including $p_{\perp} = 0$,
and therefore there is no UV/IR mixing in this case.
On the other hand, if the factor 
$e^{-\frac{p_{\perp}^2+k_{\perp}^2}{2|eB|}}$ were absent in
integrand (\ref{integr}), we would get the integral
\ba
\int \frac{d^2 k_{\perp}}{(2\pi)^2} 
e^{\frac{i}{eB}(p^1k^2-p^2k^1)}D_{\pi}^{(tree)}(k)
\label{integr1}
\ea
which diverges quadratically at $p_{\perp} = 0$, i.e., the
UV/IR mixing would occur.

Let us now turn to the description using the smeared fields.
The relation (\ref{smeared}) between the fields $\pi$ and $\Pi$
implies that their propagators are related as 
\ba
D_{\Pi}(p) =  e^{\frac{-p^{2}_{\perp}}{2|eB|}}  D_{\pi}(p).
\lb{Ppropagator}
\ea
Since $e^{\frac{-p^{2}_{\perp}}{2|eB|}}$ is an entire function,
the absence of the UV/IR mixing in the propagator $D_{\pi}$ implies
that there is no UV/IR mixing also in the propagator $D_{\Pi}$.
This conclusion
can be checked directly, by adapting the calculations of 
the one-loop correction in the propagator $D_{\pi}$
to 
the $D_{\Pi}$ propagator. In this case, it is
the form-factor $e^{\frac{-p^{2}_{\perp}}{2|eB|}}$, built in the
propagator $D_{\Pi}^{(tree)}(p)$, that is responsible for the absence 
of
the UV/IR mixing. 

This concludes the analysis in $3 + 1$ dimensions.
In the next Section, we will generalize this analysis to
arbitrary dimensions $D = d + 1$ with $d \geq 2$.

\section{The NJL model in a magnetic field
as a NCFT: Beyond 3+1 dimensions}
\label{5}

In this Section, we will generalize our analysis to
arbitrary dimensions $D = d + 1$ with $d \geq 2$.
We begin by considering the NJL model in a magnetic field
in 2+1 dimensions, choosing   
its Lagrangian density similar to that in 3 + 1 dimensions:
\ba
L = \frac{1}{2}[\bar{\psi}, (i\gamma^{\mu}D_{\mu})\psi]  +
\frac{G_2}{2N} \left[(\bar{\psi}\psi)^2 + (\bar{\psi}i\gamma^5\psi)^2 
\right].
\label{2+1action}
\ea
Here a reducible four-dimensional representation of the Dirac matrices 
is used (for details, see \cite{GMS1}). In a weak coupling
regime, the dynamical mass in this model is \cite{GMS1}
\ba
m=\frac{G_{2}|eB|}{2\pi}.
\label{2+1solution}
\ea
The LLL propagator is obtained from the (3+1)-dimensional propagator
in Eqs. (\ref{LLLpropagator}) and (\ref{fourier}) by just omitting the 
$x^3$ and $k^3$ variables there:
\ba
S(x,y)=P(\vec{x},\vec{y})S_{||}(x^0-y^0),
\label{2+1propagator}
\ea
where, instead (\ref{flatspace}), the expression for $S_{||}(x^0-y^0)$ 
is:
\ba
S_{\|}(x^0-y^0) = \int \frac{dk_0}{2\pi}
e^{ik_0(x^0-y^0)} \frac{i}{k_0\gamma^0 - m}\,
\frac{1 - i\gamma^1\gamma^2 \mbox{sign}(eB)}{2}.
\lb{2+1flatspace}
\ea
The analysis now proceeds as in the $3 + 1$ dimensional case.
The present model corresponds to a noncommutative field theory
describing neutral composites $\sigma$ and $\pi$.
Its action written
through the star product is: 
\ba
\Gamma_{2} = \frac{N|eB|}{2\pi} \int d^2x\,
\left[-iTr_{||}\left[{\cal P}\,
\mbox{Ln} \left( i\gamma^0\partial_0 -
(\Sigma + i\gamma^5\Pi)\right)\right] -
\frac{\pi}{G_2|eB|}\int dx^0({\sigma}^2+{\pi}^2)\right]_{*}\,,
\lb{2+1starACTION}
\ea
where $\Sigma$ and $\Pi$ are smeared fields (compare with 
Eq. (\ref{starACTION})). 
The action can be also written directly in the noncommutative coordinate
space: 
\ba
\Gamma_{2} = \frac{N|eB|}{2\pi} \mbox{\bf Tr}\,
\left[-iTr_{||}\left[{\cal P}\,
\mbox{Ln} \left( i\gamma^0\partial_0 -
(\hat\Sigma + i\gamma^5\hat\Pi)\right)\right] -
\frac{\pi}{G_2|eB|}\int dx^0(\hat{\sigma}^2+\hat{\pi}^2)\right]
\lb{2+1ncACTION}
\ea
(compare with Eq. (\ref{ncACTION})).

In the previous Sections, it was shown that in the
regime with the LLL dominance, the divergences in
(3 + 1)-dimensional model are generated only by the
(1+1)-dimensional longitudinal dynamics. For the
(2 + 1)-dimensional model in this regime, a stronger
statement takes place: the model is {\it finite}. It can be shown by 
repeating 
the analysis used in 3 + 1 dimensions. In particular, 
in the continuum limit $\Lambda \to \infty$,
the effective potential in this model is finite without
any renormalizations:
\ba
V_{2}(\sigma,\pi) = \frac{N(\sigma^2 + \pi^2)}{2G_2} -
\frac{N|eB|\sqrt{\sigma^2 + \pi^2}}{2\pi}.
\lb{2+1pot}
\ea
Using Eq. (\ref{2+1solution}), one can express the coupling constant 
$G_2$ in the potential through $m$ and $|eB|$. Then the potential 
takes an especially simple form:
\ba
V_{2}(\sigma,\pi) = \frac{N|eB|}{2\pi}\left(\frac{\sigma^2 + \pi^2}{2m} -
\sqrt{\sigma^2 + \pi^2}\right).
\lb{2+1pot1}
\ea

For momenta $k \ll \sqrt{|eB|}$,
the dispersion relation for $\pi$ excitations is \cite{GMS1}
\ba
E_\pi \simeq \frac{\sqrt{2}m}{{|eB|}^{1/2}}(\vec{k}^2)^{1/2}.  
\lb{2+1dispersion}
\ea
Therefore, as in 3+1 dimensions, the velocity  
$|{\vec v}|=|\partial E_{\pi}/\partial {\vec k}|$
is strongly suppressed: in the present case it is of
order $m/{|eB|}^{1/2}$.
As to the $\sigma$ excitation, its "mass", defined as the energy 
at zero momentum, is very large: $M_{\sigma} \sim 
(\sqrt{eB}/m)^{1/2}\,\sqrt{|eB|}$ \cite{GMS1}. Therefore the $\sigma$-mode 
decouples from the dynamics with $k \ll \sqrt{eB}$. 

As in the case of 3+1 dimensions,  
this truncated
[based on the LLL dynamics] model is self-contained and
one can consider arbitrary large values for momenta
there. It is easy
to check that for $k \gg \sqrt{|eB|}$
the dispersion relation for $\pi$ excitations takes the
form
\ba
E_{\pi} \simeq m\left(2-
e^{-\frac{\vec{k}^2}{2|eB|}}\right).
\lb{dispersion2+1}
\ea
In this regime, 
the velocity becomes extremely small, 
$|\vec{v}| \sim \frac{m|\vec{k}|}{|eB|}
e^{-\vec{k}^2/2|eB|}$, and a $\pi$ excitation is a loosely bound
state. Its mass is close to the $2m$ threshold.

As was shown in Section \ref{4}, in the 
limit $|eB| \to \infty$ the (3+1)-dimensional model is reduced
to a continuum set of independent
(1+1)--dimensional Gross--Neveu models labeled by the
coordinates in the plane perpendicular to the magnetic field.
Similarly to that, in the case of 2 + 1 dimensions, in the limit
$|eB| \to \infty$ the model is reduced to a set of (0+1)-
dimensional (i.e., quantum mechanical) models labeled by
two spatial coordinates.

A new feature of the (2+1)-dimensional
model is a confinement dynamics for charged
particles: they do not propagate
in a magnetic background.
On the other
hand, since neutral composites are free to propagate in a magnetic 
field,
one can define asymptotic states and 
$S$-matrix for them. The $S$-matrix should be unitary in the
subspace of neutral composites.
 
Let us now consider the case of higher dimensions
$D=d+1$ with $d>3$. First of all, recall 
that for an even $d$, by using spatial rotations, 
the noncommutativity tensor $\theta^{ab}$ in a noncommutative theory
with $[\hat{x}^a,\hat{x}^b]=i\theta^{ab}$
can be reduced to the following canonical skew-diagonal
form with skew-eigenvalues $\theta^a,\, a=1,...,\frac{d}{2}$ \cite{DNS}:
\ba
\theta^{ab} = \left(%
\begin{array}{ccccccccc}
  0 & \theta^1 & & & & & & & \\
  -\theta^1 & 0 & & & & & & & \\
  & & & \cdot & & & & & \\
  & & & & \cdot & & & & \\
  & & & & & \cdot & & & \\
  & & & & & & & 0 & \theta^{d/2}  \\
  & & & & & & & -\theta^{d/2} & 0
\end{array}%
\right).
\label{canonical}
\ea
If $d$ is odd, then the number of canonical skew-eigenvalues of 
$\theta^{ab}$
is equal to $[\frac{d}{2}]$, where $[\frac{d}{2}]$ is the integer part 
of
$d/2$, and the canonical form of $\theta^{ab}$ is similar to 
(\ref{canonical})
except that there are additional one zero column and one
zero row.

On the other hand, a constant magnetic field in $d$ dimensions is 
also characterized by
$[\frac{d}{2}]$ independent parameters, and the strength tensor
$F^{ab}$ can be also reduced to the canonical skew-diagonal
form \cite{Gr,Gt}:
$$
F^{ab} = \sum_{c=1}^{[\frac{d}{2}]} B^{c} (\delta^a_{2c-1}\delta^b_{2c} 
-
\delta^b_{2c-1}\delta^a_{2c}).
$$
The corresponding nonzero components of the vector potential are equal to
$$
\vec{A}^{ext} = (-\frac{B^1x^2}{2},\,\frac{B^1x^1}{2},\, ... \,,
-\frac{B^{[\frac{d}{2}]}x^{2[\frac{d}{2}]}}{2},\,
\frac{B^{[\frac{d}{2}]}x^{2[\frac{d}{2}]-1}}{2}).
$$
Thus, we see that there is one-to-one mapping
between the 
skew-eigenvalues 
of the noncommutativity tensor $\theta^{ab}$ and the independent
parameters of the spatial part of the electromagnetic strength tensor
$F^{ab}$
in a space of any dimension $d \geq 2$.

Chiral symmetry breaking in the NJL model in a strong magnetic field
in dimensions with $d>3$ was studied in
\cite{Gr}. By using results of that paper, it is not difficult
to extend our analysis in 3+1 and 2+1 dimensions to the case
of $d > 3$.
The crucial point in the analysis is the structure of
the Fourier transform of the translationally invariant part
of the LLL propagator. If all $B^a$ are nonzero, 
one can show that it is 
\ba
\tilde{S}_{[\frac{d}{2}]}(k) = 
i\, \exp\left[-\sum_{a=1}^{[\frac{d}{2}]}\frac{k_{2a-1}^2 +\,
k_{2a}^2}{|eB^a|}\right]\,
\frac{k_{||}\gamma^{||} + m}{k_{||}^2-m^2}\, \Pi_{a=1}^{[\frac{d}{2}]}
(1-i\gamma^{2a-1}\gamma^{2a}\mbox{sign}(eB^a)),
\lb{d+1fourier}
\ea
where $k^{||} = k^0$ if $d$ is even and $k^{||} = (k^0,k^d)$ if $d$ is
odd. If some $B^c=0$, then, for each $c$, 
the longitudinal part $k^{||}$ gets two additional components, 
$k^{2c-1}$ and $k^{2c}$, and the corresponding terms are absent 
in the transverse part of expression (\ref{d+1fourier}).
Thus, like in 3+1 and 2+1 dimensions,
the LLL propagator factorizes into the transverse and longitudinal
parts. The projection operator
$P_{n}(x_{\perp},y_{\perp})$ on the LLL
is now equal to the direct product of the projection 
operators (\ref{projector})
in the $x^{2a-1}x^{2a}$-planes with nonzero $B^{a}$
[here the subscript $n$ is the number of nonzero independent 
components of $F^{ab}$]. 

Because of that, it is clear that the NJL model in a strong 
magnetic field
in a space of arbitrary dimensions $d \geq 2$ 
corresponds to
a noncommutative field theory with parameters
$\theta^{ab}$ expressed through the magnetic part of the 
strength tensor $F^{ab}$. Its action is
[compare with expressions (\ref{ncACTION}) and (\ref{2+1ncACTION})]:
\ba
\Gamma_{n} = N \mbox{\bf Tr}\,\left[
-\frac{i\Pi_{a=1}^{n}|eB^a|}{(2\pi)^{n}}
\, Tr_{||}\left[{\cal P}_n\,
\mbox{Ln} \left( i\gamma^{||}\partial_{||} -
(\hat{\Sigma} + i\gamma^5\hat{\Pi})\right)\right] -
\frac{1}{2G_d}\int d^{D-2n}x_{||}\,(\hat{\sigma}^2+\hat{\pi}^2)
\right]\,,
\lb{d+1ncACTION}
\ea
where $n$ is the number of nonzero independent components of $F^{ab}$
and the projector ${\cal P}_n$ equals the direct product
of projectors (\ref{P_}) in the $x^{2a-1}x^{2a}$-planes with nonzero 
$B^{a}$.
In particular, 
for a magnetic field configuration with the maximal number 
$n = [d/2]$
of independent nonzero tensor components,
the dynamics is quasi-(1+1)-dimensional for
odd $d$ and finite for even $d$. 
In the latter case the model
describes a confinement dynamics of charged particles. Also,
as all $|eB^a| \to \infty$, the model is reduced either to a continuum 
set of
(1+1)-dimensional GN models labeled by $d-1$ spatial coordinates
(odd $d$) or to a set of quantum mechanical models labeled by
$d$ spatial coordinates (even $d$).

\section{Conclusion}
\label{6}

The main result of this paper is that in any dimension $D=d+1$ 
with $d \geq 2$,
the NJL model in a strong magnetic field 
determines a consistent NCFT. These NCFT are
quite sophisticated that is reflected in 
their action (\ref{d+1ncACTION}) expressed
through the smeared fields $\Sigma$ and $\Pi$ with
built-in exponentially damping form-factors.
These 
form-factors occur in the propagators of the smeared fields
and are responsible for removing the UV/IR mixing that plagues 
conventional nonsupersymmetric NCFT \cite{MRS}. As an alternative,
one can also use
the composites fields $\sigma$ and $\pi$. 
In this case, the form-factors occur in their
interaction vertices and this again leads to the removal of the
UV/IR mixing.

An especially interesting case is that for
a magnetic field configuration with the maximal number $[d/2]$
of independent nonzero tensor components. In that case,
the dynamics is quasi-(1+1)-dimensional for
odd $d$ and finite for even $d$.
How can it be, despite
the fact that the initial NJL model is nonrenormalizable
for $d \geq 2$? And, moreover, how can it happen in theories
in which neutral composites propagate in a bulk of
a space of {\it arbitrary} high dimensions? The answer to these 
questions is straightforward. 
The initial NJL
model in a strong magnetic field and the truncated model
based on the LLL dynamics are essentially identical only
in infrared, with momenta $k \ll \sqrt{|eB|}$. At
large momenta, $k \gg \sqrt{|eB|}$, these two models
are very different. 
It is the LLL dominance
that provides the exponentially damping
(form-)factors which are responsible for finiteness
of the present model for even $d$ and its 
quasi-(1+1)-dimensional character for odd $d$.  
Thus, besides being a low
energy theory of the NJL model in a strong magnetic field, the
NCFT based on the LLL dynamics is self-contained and
self-consistent.

As was discussed in Section \ref{2}, the exponentially damping factors
occur also in nonrelativistic quantum mechanical models. In
particular, they are an important ingredient of the formalism
of the projection onto the LLL developed for studies of 
condensed matter systems in Refs. \cite{GK,DJ}.
It is then natural to ask why do such factors not appear 
also in string theories in a magnetic field? 
The answer to 
this question
is connected with a completely different way that open strings respond 
to a strong $B$ field. It can be seen already on the classical 
level. Indeed, due to the boundary
conditions at the ends of open strings, their length {\it grows} with 
$B$ until the string tension compensates the 
Lorentz forces exerted at the
ends of strings \cite{Sheikh}. In contrast to that, 
in quantum field and condensed matter systems, 
charged particles, 
which form neutral composites,
move along circular orbits in a magnetic field, and their
radius {\it shrinks} with increasing $B$. This leads to the Landau 
type wave functions 
of composites and, therefore, to the  
exponentially damping (form-)factors either in vertices
(for $\sigma$ and $\pi$ fields) or in propagators (for
smeared fields). 

Therefore, unlike the dynamics
of neutral composites in
condensed matter and quantum field systems, open strings in a 
magnetic background do lead to the conventional NCFT.
Since these theories are supersymmetric, the UV/IR mixing 
affects only the constants of renormalizations and does not
destroy their consistency \cite{super}.
Thus different physical systems in a magnetic
fields lead to different classes of consistent NCFT.
\footnote {In this regard, quantum mechanical systems in
a strong magnetic field are special. As was shown in
Section \ref{2}, depending on two different treatments of
the case with the mass $m=0$, they determine either
conventional NCFT, as it is done in Ref. \cite{BS}, or
NCFT with exponentially damping form-factors.}  

In the present paper, as an example, we considered the NJL model
in a magnetic field.
It is however clear that because of the universality of
the dynamics connected with transverse coordinates, 
form-factors should also occur in propagators
(or in vertices) in more complicated field theories in a magnetic
field (although the form of the form-factors can vary). 
Therefore the
corresponding NCFT should be in this regard similar to those 
revealed in this paper.

\begin{acknowledgments}
We thank I. Shovkovy for useful discussions.
The work was supported by the Natural 
Sciences and Engineering Research Council of Canada.
V.A.M. thanks the Institute for Nuclear Theory at the University
of Washington for its hospitality and Department of Energy for
partial support during the completion of this work.

\end{acknowledgments}

\vspace{8mm}

\centerline{\bf APPENDIX A}

\vspace{8mm}

In this Appendix, we will show that in the case
of fields independent of the longitudinal coordinates
$x_{||}$, all their interaction
vertices $\Gamma_{n\phi}$ ($n \geq 3$)  
can be rewritten through
the star product. 

The relevant part
of the n-point vertex $\Gamma_{n\phi}$ is the part which includes the
integration
over transverse coordinates. It has the form:
\ba
\Gamma^{\perp}_{n\phi} \equiv \int d^2x_1 ... d^2x_{n}
P(x_1,x_2)\phi(x_2)P(x_2,x_3)\phi(x_3) ...
P(x_n,x_1)\phi(x_1),
\label{perp1}
\ea
where $P(x_1,x_2)$ is the transverse part of the fermion propagator
written in Eq. (\ref{projector})
[here, for convenience, we omitted the subscript $\perp$ in
transverse coordinates].

Expressing the fields
$\phi$ through their Fourier transforms, one can explicitly integrate
over $x_i$ coordinates in (\ref{perp1})
[the integrals are Gaussian].
It can be done step by step. First, we find
\ba
I_1(x_1,x_3) = \int d^2x_2 P(x_1,x_2) e^{i\vec{k}_2\vec{x}_2} P(x_2,x_3)
= P(x_1,x_3) e^{-\frac{\vec{k}_{2}^2}{2|eB|}}
e^{\frac{sign(eB)\epsilon^{ab}k_2^a(x_1-x_3)^b}{2}}
e^{\frac{i}{2}\vec{k}_2(\vec{x}_1 + \vec{x}_3)}.
\label{I1}
\ea
The second step leads to an expression with a similar structure:
$$
I_2(x_1,x_4) = \int d^2x_3 I_1(x_1,x_3) e^{i\vec{k}_3\vec{x}_3} P(x_3,x_4)
=
$$
\ba
P(x_1,x_4) e^{-\frac{\vec{k}_2^2+\vec{k}_3^2+\vec{k}_2
\vec{k}_3}{2|eB|}}
e^{-\frac{i}{2eB}\epsilon^{ab}k_2^ak_3^b} 
e^{\frac{sign(eB)\epsilon^{ab}
(k_2+k_3)^a(x_1-x_4)^b}{2}}
e^{\frac{i}{2}(\vec{k}_2+\vec{k}_3)(\vec{x}_1+\vec{x}_4)}.
\label{I2}
\ea
Proceeding in this way
until the integration over $x_n$, we encounter the
integral
\ba
I_{n-1}(x_1,x_1) = \int d^2x_n
I_{n-2}(x_1,x_n) e^{i\vec{k}_n
\vec{x}_n}P(x_n,x_1).
\label{Ilast}
\ea
It closes the fermion loop because the
last argument in $P(x_n,x_1)$ coincides with the first
argument of $I_{n-2}$. Because of that,
the result of this integration is especially simple:
\ba
I_{n-1}(x_1,x_1) = \frac{|eB|}{2\pi}
e^{-\frac{\sum_{i=2}^n\vec{k}_i^2+\sum_{2 \le i < j}^n
\vec{k}_i\vec{k}_j}{2|eB|}}\,\,\,
e^{-\frac{i}{2eB}(\sum_{2 \le i < j}^n\epsilon^{ab}k_i^ak_j^b)}\,\,\,
e^{i(\sum_{i=2}^{n}\vec{k}_i\vec{x}_1)},
\label{int}
\ea
where the equality $P(x_1,x_1)= |eB|/2\pi$ was used. The last integration
over $x_1$ yields
\ba
I_n = \int d^2x_1 I_{n-1}(x_1,x_1) e^{i\vec{k}_1\vec{x}_1} =
2\pi|eB|\delta^2(\sum_{i=1}^n \vec{k}_i)\,\,
e^{-\frac{\sum_{i=2}^n\vec{k}_i^2+\sum_{2 \le i < j}^n
\vec{k}_i\vec{k}_j}{2|eB|}}\,\,\,
e^{-\frac{i}{2eB}(\sum_{2 \le i < j}^n\epsilon^{ab}k_i^ak_j^b)}.
\label{int1}
\ea
Here the delta function ensures the conservation of the
total transverse momentum.
Now, because of the identity
$$
\sum_{i=2}^n\vec{k}_i^2+\sum_{2 \le i < j}^n \vec{k}_i\vec{k}_j =
-\sum_{1 \le i < j}^n\vec{k}_i\vec{k}_j + (\sum_{i=1}^n \vec{k}_i)^2
-
\vec{k}_1(\sum_{i=1}^n\vec{k}_i)
$$
and the conservation of the total momentum, we obtain the equalities
$$
\sum_{i=2}^n\vec{k}_i^2+\sum_{2 \le i < j}^n \vec{k}_i\vec{k}_j =
-\sum_{1 \le i < j}^n\vec{k}_i\vec{k}_j =
\frac{1}{2}\sum_{i=1}^n\vec{k}_i^2
$$
and
$\sum_{2 \le i < j}^n\epsilon^{ab}k_i^ak_j^b =
\sum_{1 \le i < j}^n\epsilon^{ab}k_i^ak_j^b$.
Using these equalities,
we conclude that the
exponential term in expression 
(\ref{int1}) can be rewritten through the cross
product as
$e^{-\frac{\sum_{i=1}^n\vec{k}_i^2}{4|eB|}}\,e^{-\frac{i}{2}
\sum_{i<j} k_i \times k_j}$ . Therefore, similarly to three and four point 
vertices (\ref{3point1}) and (\ref{4point}),
a generic n-point interaction
vertex $\Gamma_{n\Phi}$ ($n \geq 3$) has the following structure:
\ba
\Gamma_{n\Phi} =  C_{n}\frac{N|eB|}{m^{n-2}}
\int d^2x_{||}\frac{d^2k_1...d^2k_n}{(2\pi)^{2n}}
\Phi(k_1)... \Phi(k_n) \delta^2(\sum_i k_i)
e^{-\frac{i}{2} \sum_{i<j} k_i \times k_j},
\label{vertexA}
\ea
where here $\Phi$ represents the smeared fields $\Pi$ and
$\tilde{\Sigma}$ and $C_n$ is a numerical constant which can be easily
found by expanding the effective potential in the Taylor series
in constant fields $\pi$ and $\tilde{\sigma}$.
Equation (\ref{vertexA}) in turn
implies that the
vertex $\Gamma_{n\Phi}$ 
can be rewritten through the star product
in the coordinate space as
\ba
\Gamma_{n\Phi} =  C_{n}\frac{N|eB|}{4\pi^{2}m^{n-2}}
\int d^2x_{||}d^2x_{\perp}\,\,\,
\Phi_1 * \Phi_2 *... * \Phi_n
\label{vertexA1}
\ea
(compare with expressions in Eq. (\ref{xspacevertices})).
In the noncommutative coordinate space, the vertex is:
\ba
\Gamma_{n\Phi} =  C_{n}\frac{N|eB|}{4\pi^{2}m^{n-2}}
\mbox{\bf Tr}\,
\hat{\Phi}_1\hat{\Phi}_2...\hat{\Phi}_n
\label{vertexA2}
\ea
(compare with Eq. (\ref{ncspacevertices})).

\centerline{\bf APPENDIX B}

\vspace{8mm}

In this Appendix, it will be shown that the
exponentially damping factors and
the $M$-star product
naturally appear in the formalism of the projected density operators
on the LLL states developed
in studies of the
quantum Hall effect in Ref. \cite{Sinova}. 
To be concrete,
we will consider the $\Gamma_{4\pi}$ vertex in this formalism.

As follows from Eq. (\ref{action}), the
$\Gamma_{4\pi}$ vertex is given
by
\ba
\Gamma_{4\pi} = \frac{i}{4} \int d^4x d^4y d^4z d^4v\,\,\, \mbox{tr}
\left[ S(x,y)\gamma^5 \pi(y) S(y,z) \gamma^5 \pi(z) S(z,v)
\gamma^5 \pi(v) S(v,x) \gamma^5 \pi(x) \right].
\label{A1}
\ea
According to Eq. (\ref{factorization}), the dependence on the transverse
$x_{\perp}$ and longitudinal $x_{\|}$ coordinates factorizes in the
LLL propagator $S(x,y)$. If
fields $\pi$ in (\ref{A1}) do not depend on $x_{\|}$, then it is
straightforward to integrate over the longitudinal coordinates in
this expression that yields the factor
\ba
\frac{i}{4} \int \frac{d^2k_{\|}}{(2\pi)^2}\,\,\, \mbox{tr}\,\,\,
\left( \frac{1}{k_{\|}\gamma^{\|} - m}
\frac{1-i\gamma^1\gamma^2\,\mbox{sign}(eB)}{2} \gamma^5 \right)^4 =
-\frac{1}{8\pi m^2}.
\label{longitudinal}
\ea
To get the $\Gamma_{4\pi}$ vertex, we now need to calculate
the transverse part
\ba
\int d^2x_{\perp}d^2y_{\perp}d^2z_{\perp}d^2v_{\perp}
P(x_{\perp},y_{\perp})
\pi(y_{\perp}) P(y_{\perp},z_{\perp})
\pi(z_{\perp}) P(z_{\perp},v_{\perp}) \pi(v_{\perp})
P(v_{\perp},x_{\perp}) \pi(x_{\perp}).
\label{transverse}
\ea
We will use the formalism of projected density
operators \cite{Sinova} to calculate it.
The crucial point is the fact that the
transverse part of the LLL fermion propagator $P(x,y)$
is the the projection operator on the LLL states (henceforth
we will omit the subscript $\perp$ for the transverse
coordinates). Namely,
\ba
P(x,y) = \sum_n <x|n><n|y>,
\label{Aprojector}
\ea
where the sum is taken over all LLL states, which in the symmetric
gauge are
\ba
\psi_n(z,z^*) = \left(\frac{|eB|}{2}\right)^{\frac{n+1}{2}}  
\frac{z^n}{\sqrt{\pi n!}}
e^{-\frac{|eB|zz^*}{4}}
\label{wf}
\ea
with $z=x^1- i\,\mbox{sign}(eB)x^2$. Now, by using completeness relations 
like
$$
\int d^2y <n_1|y> \pi(y) <y|n_2> = <n_1| \pi |n_2>,
$$
we obtain expression (\ref{transverse}) in the form
\ba
\sum_{n_1,...,n_4} <n_1|\pi|n_2>...<n_3|\pi|n_4>.
\label{A3}
\ea
To get the $\Gamma_{4\pi}$ interaction vertex in the momentum space, we
will use the Fourier transforms of fields $\pi$. Then we encounter
factors of the form:
$$
<n_i|\rho_k|n_j>,
$$
where $\rho_k=e^{i\vec{k}\vec{x}}=\exp\left[\frac{i}{2}
(kz^* + k^*z)\right]$, with $k=k^1 - i\,\mbox{sign}(eB)k^2$, 
is called the density operator. 

In what follows, we will use the methods developed in
Refs. \cite{GK,DJ} and, in
fact, follow very closely Ref. \cite{GZ}.
First of all,
since the prefactor in expression (\ref{wf}) is analytic in $z$,
the
factor $e^{\frac{i}{2}k^*z}$ in $\rho_k$
acts entirely within the LLL. On the other hand, another
factor $e^{\frac{i}{2}kz^*}$ in $\rho_k$ contains $z^*$
and therefore does not act within the LLL.
Actually, the following relation takes place:
\ba
<n|(z^*)^s|m> = <n|\left( \frac{2}{|eB|}\frac{\partial}{\partial z}
+ \frac{z^*}{2} \right)^s|m>,
\label{projection}
\ea
which expresses the matrix elements of $z^*$ between
the LLL states
in terms of the
operator
$\hat{z} = \frac{2}{|eB|}\frac{\partial}{\partial z} +
\frac{z^*}{2}$ that acts within the LLL.
Therefore, on the LLL states,
we can replace the density operator $\exp\left[\frac{i}{2}(kz^*
+ k^*z)\right]$ by the projected density operator
$\hat{\rho}_k = e^{\frac{i}{2}k\hat{z}}e^{\frac{i}{2}k^*z}$.

Now, using the the projected density operators $\hat{\rho}_k$
for the $\Gamma_{4\pi}$ vertex, we get:
\ba
\Gamma_{4\pi}=-\frac{1}{8\pi m^2} \int d^2x_{||}\int
\frac{d^2k_1 d^2k_2 d^2k_3 d^2k_4}{(2\pi)^8}\pi(k_1)\pi(k_2)\pi(k_3)\pi(k_4)
\sum_{n_1,...,n_4}(\hat{\rho}_{k_1})_{n_1n_2}(\hat{\rho}_{k_2})_{n_2n_3}
(\hat{\rho}_{k_3})_{n_3n_4}(\hat{\rho}_{k_4})_{n_4n_1}.
\label{A4}
\ea
Since the LLL states form a complete basis for 
the operators $\hat{\rho}_k$, we have
$$
\sum_{n_2} (\hat{\rho}_{k_1})_{n_1n_2} (\hat{\rho}_{k_2})_{n_2n_3} =
(\hat{\rho}_{k_1}\hat{\rho}_{k_2})_{n_1n_3}.
$$
The product of two projected density operators is given by
\cite{GZ}
\ba
\hat{\rho}_{k_1} \hat{\rho}_{k_2} = \exp \left[ 
\frac{\vec{k}_1\vec{k}_2}{2|eB|} - 
\frac{i}{2}k_1 \times k_2
\right] \hat{\rho}_{k_1+k_2}.
\label{product}
\ea
Notice that the exponent in this equation can be rewritten
through the $M$-cross product (\ref{Mcross}) as
$e^{-\frac{i}{2} k_1 \times_M k_2}$.

Therefore, we find the following expression for $\Gamma_{4\pi}$:
\ba
\Gamma_{4\pi} = -\frac{1}{8\pi m^2} \int d^2 x_{||} \int
\frac{d^2k_1 d^2k_2 d^2k_3 d^2k_4}{(2\pi)^8} \pi(k_1)\pi(k_2)\pi(k_3)\pi(k_4)
\,e^{-\frac{i}{2}\sum_{i<j} k_i \times_M k_j}
\sum_{n} (\hat{\rho}_{k_1+k_2+k_3+k_4})_{nn}.
\label{A5}
\ea
Using further the relation (see Ref. \cite{Sinova})
$$
\sum_{n} (\hat{\rho}_{k_1+k_2+k_3+k_4})_{nn} = N \delta_{\sum_i\vec{k}_i,0}\,,
$$
where $N = S\, \frac{|eB|}{2\pi}$ is the number of states on
the LLL and $S$ is the square of the transverse plane, and
the identity
$$
S\, \delta_{\sum_i\vec{k}_i,0} = (2\pi)^2\delta^2(\sum_i\vec{k}_i),
$$
we finally get the expression for the vertex
$\Gamma_{4\pi}$ that
coincides with
expression (\ref{xspacevertices1}).

Thus we see that
the mathematical reason for
the appearance of exponentially damping factors and the $M$-star product
is related to the algebra of the projected density operators (\ref{product}).
Obviously, the generalization of the above calculations to an
arbitrary interaction vertex for the
$\pi$ and $\tilde{\sigma}$ fields is straightforward.

\end{document}